
\magnification=1200

\baselineskip=17pt 

\nopagenumbers


\voffset=1truecm 
\hsize= 5.25in 
\vsize=7.5in 

\def\({\c c }

\def\|{\'\i}

\def\nl{\par\noindent}

\vskip 1.5truecm

\pretolerance=10000    

\centerline {\bf Light-front Quantized Scalar Field Theory and
Phase Transition } \par

\bigskip
\bigskip
\bigskip
\centerline {Prem P. Srivastava\footnote{*}{{\it Postal address: Rua Xavier
Sigaud 150, 22290 Rio de Janeiro, RJ, Brasil.}
\nl {\bf e-mail:} prem@cbpfsu1.cat.cbpf.br}}
\medskip
\nl \centerline {\it Centro Brasileiro de Pesquisas F\|sicas, Rio de Janeiro}
\nl\centerline {\it and }
\nl\centerline {{\it INFN {\sl and} Dipartimento de Fisica} {\sl Galileo
Galilei},
{\it Universit\`a di Padova, Padova.}}

\bigskip
\bigskip
\vskip 0.5cm

\centerline  {\bf Abstract:}
\bigskip
\medskip
{\leftskip 36pt \rightskip 36pt {\quad The
light-front Hamiltonian formulation for the scalar field theory
contains a new ingredient in the form of a constraint equation.
Renormalization of the two dimensional $\phi^{4}$ theory,
described in the continuum,
is discussed. The mass renormalization condition and the renormalized
constraint equation contain all the information
to describe the phase transition in the theory, which is found to be
of the second order. We argue that the same
result would also be obtained in the conventional equal-time formulation.
  }\par}

\vfill \eject

\nopagenumbers

\headline={\rightheadline}
\def\rightheadline{\tenrm\hfil\folio} 
\pageno=1

\baselineskip=17pt

\nl {\bf 1- Introduction:}

\medskip
Dirac [1] suggested the possibility of building
dynamical theory of a physical system on
the three dimensional hypersurface in space-time formed by a plane wave
front advancing with the velocity of light. The initial conditions on the
dynamical variables are here specified
on the hyperplane ({\it light-front}), say, $\,\tau=(x^{0}+x^{3})/
{\sqrt 2}=0,\,$ which has a light like normal- the {\it front form } dynamics.
The  conventional formulation uses
instead the $x^{0}\equiv t=0$
hyperplane- the {\it instant form} dynamics.  In {\it both the cases} the
separation between two points is space like in general which may become
light like in particular cases.
Latter
the {\it front form} dynamics was rediscovered by Weinberg [2] in the
infinite momentum frame rules in the
quantized field theory. The rules  were clarified by Kogut and Soper [3]
to correspond  to the theory quantized on the light-front. Even earlier [4]
the ${ p\to\infty}$ technique played an important role in the derivation of
the current algebra sum rules and it was also observed that it amounted to
using appropriate light-front current commutators.
The recent revival of interest [5,6] in the light-front
theory has been motivated
by the difficulties faced in the nonperturbative QCD in
the conventional  formulation.
We have  the  problem of reconciling
the standard constituent (valence) quark model and QCD, where the vacuum state
itself
contains an infinite sea  of constituent quarks and gluons ( partons), to
describe the hadrons.  The problem of describing
relativistic bound-state of light quarks
in the presence of the complicated vacuum in the {\it instant form}
seems difficult but it was found that the
Light-front Tamm-Dancoff method [5] may be feasible.
The {\it front form} dynamics may serve as a
complementary tool where we have a simple vacuum while the
complexity of the problem is now transfered to the
light-front Hamiltonian. This was illustrated recently also
in the light-front quantized scalar field theory  where the
light-front Hamiltonian  corresponding to the
to the polynomial form of the equal-time formulation is found [7] to be
nonlocal.
This follows
due to the presence [8,7] of a {\it constraint equation}. It was
shown  [7] that the  physical outcome as regards the spontaneous symmetry
breaking in the case of continuous symmetry
in $3+1$ dimensions
is the same in the
two forms of dynamics, even though
achieved  through different mechanisms.
In fact, we will show
below  that many of the external ingredients,
which we usually {\it add} to constrain the scalar theory treated in the
{\it instant form } upon invoking the physical considerations, are already
contained in the {\it front form} through a set of
self-consistency  constraints.

The simplicity of the
light-front vacuum, which may often coincide with the free theory
one, derives in particular from  the observation  that
the longitudinal momentum  $k^{+}=(k^{0}{+}k^{3})/{\sqrt 2}\;$ of
a  free massive particle is necessarily positive. Another
general feature of the {\it front form} theory is that it describes
a constrained dynamical system. The Dirac procedure [9]
to construct a self-consistent Hamiltonian formulation is
convenient to use. It allows us also to
unify [1] the principle of
(special) relativity in the theory.
We also note that  if we assume the {\it microcausality}
principle, both the equal-time  and
equal-$\tau$ commutators of two, say,  scalar
observables may take nonvanishing values only on the
light-cone.

We  discuss in the paper
the  stability of the vacuum in the $\phi^{4}$ theory
when the coupling constant is increased from vanishingly small values to
larger values. The light-front framework seems very attractive to study
this problem. On renormalizing the theory we have now,
in addition to the usual equations like the mass renormalization
conditions, also the  renormalized constraint
equations to deal with. In our context we remind
that there are rigorous proofs [10] on the triviality of
$\phi^{4}$ theory in the continuum for more than four space-time
dimensions and on its interactive nature for dimensions less than four.
In the important case of four dimensions the
situation is still unclear [10]  and light-front dynamics
may throw some light on it. In view of the
complexity of the renormalization problem in this case
we will illustrate our points by
considering  only the two dimensional theory, which is of importance
in the condensed matter physics.  For example, from the well established
results on the generalized Ising models, Simon and Griffiths [11]
conjectured some time ago that the two dimensional $\phi^{4}$ theory should
show the  {\it second order} phase transition. We do find it to be so
by quantizing the theory, considered directly in the continuum formulation,
on the light-front. The variational
methods based on the {\it instant form }
like the Hartree approximation or Gaussian effective potential
[12], or the one based on a scheme of canonical transformations [13],
or  using the {\it front form} theory but ignoring the constraint [14],
all seem to give  a first order
phase transition contradicting the conjecture. The {\it instant form}
post-Gaussian approximation [15] and the non-Gaussian variational
method [16] give a second order transition for a
particular value of the coupling constant.
Our result shows  second order transition for any coupling above a critical
value. In view of the remarks made at  the end of the first  paragraph {\it
front form} theory should be able to throw light on this quite old
problem.
The procedure used in the paper is
the well established Dyson-Wick expansion [17] of perturbation theory and
may be improved systematically computing still higher order corrections
which is difficult to do in the variational methods.
A brief sketch of the {\it front form} Hamiltonian formulation is given in Sec.
2,
the renormalization and the phase transition are discussed in Sec. 3, and
the conclusions along with a short discussion summarized  in Sec. 4.

\bigskip
\nl {\bf 2- Light-front Hamiltonian formulation:}
\medskip

We consider the scalar field theory
with the Lagrangian density in the {\it front form}
given by
$\, [{\dot\phi}{\phi^\prime}-V(\phi)].\,$  Here an overdot and
a prime indicate the partial derivatives with respect to the light-front
{\sl time} $\,\tau\equiv x^{+}=(x^0+x^1)/{\sqrt2}\,$
and the longitudinal coordinate
$\,x \equiv x^{-}=(x^0-x^1)/{\sqrt2}\,$ respectively. The eq. of motion
is $\,2\dot{\phi'}= -V'(\phi)$, where a prime on $V$ indicates
the variational derivative with respect to $\phi$. It
shows that the classical solutions, $\phi=const.$
are allowed and given by
solving $V'(\phi)=0$.  Based on physical considerations we
write [7] $\;\phi(x,\tau)=
\omega(\tau)+\varphi(x,\tau)\;$ where the variable $\omega$ corresponds
to the bosonic condensate and $\varphi$ describes the (quantum)
fluctuations above
the former.  We would discuss the
possibility of  phase transition in the theory,
{\it described directly in the continuum}, by
studying the properties of the ground state.
Since   $\,\langle vac \vert\phi\vert vac
\rangle=\omega\,$ and the vacuum is required  to be
translational invariant (special relativity),
 it follows that the  $\omega $ must be
independent of $\tau$. In the context of the present discussion
we would then treat $\omega$ as a non-dynamical variable,
 like we frequently do in the context of equal-time formulation, and its
value will serve to characterize the vacuum state. We remark
that in the self-consistent Hamiltonian formulation constructed [7] directly
in the continuum, the variable $\omega $ is found to
commute  with the nonzero mode
operators contained in $\varphi$ and is consequently a c-number.

 The light-front Hamiltonian is found to be ($L\to\infty$)

$$ H^{l.f.}=\int_{-L/2}^{L/2} dx \,
\Bigl [\omega(\lambda\omega^2+{m_{0}}^2)\varphi+
{1\over 2}(3\lambda\omega^2+{m_{0}}^2)\varphi^2+
\lambda\omega\varphi^3+{\lambda\over 4}\varphi^4
+const.\Bigr ]. \,\eqno(2.1)$$

\nl Here $\,V(\phi)=(1/2){m_{0}}^2 {\phi^2}
+({\lambda/4})\phi^{4}+const.\,$,  $\lambda> 0$, which has the {\it
correct sign}
for the {\it bare} mass term,
and the field  $\varphi$ satisfies the well known
light-front commutator $\; [\varphi(x,\tau),\varphi(y,\tau)]=
-(i/4)\epsilon(x-y)$.

{}From the Hamilton's eq.

$$\eqalign {\dot\varphi(x,\tau)&= -i\, [\varphi(x,\tau),H^{l.f.}(\tau)]\cr
&= -{1\over 4}\,\int dy \,\epsilon(x-y)\, V'(\phi(y,\tau))}\eqno(2.2) $$

\nl we recover the Euler-Lagrange eq.

$$\dot\varphi'(x,\tau)=-{1\over 2}\, V'( \phi(x,\tau)).\,\eqno(2.3)$$

\nl assuring us of the self-consistency [9].
Substituting  the value of $V'(\phi)$ obtained
from (2.3) into (2.2) we find after an integration by parts

$$ \dot\varphi(x,\tau)=\dot\varphi(x,\tau)-
{1\over 2}\,{ \Bigl [\dot\varphi(\infty,\tau)\epsilon(\infty-x)
-\dot\varphi(-\infty,\tau)\epsilon(-\infty-x)\Bigr].}\, \eqno(2.4) $$

\nl Considering now the finite values of $x$ this leads to
$\,\dot\varphi(\infty,\tau)+\dot\varphi(-\infty,\tau)=0\,$. The
light-front commutator, on the other hand,
may be realized in the  momentum space through the expansion

$$\varphi(x,\tau)= {1\over {\sqrt{2\pi}}}\int_{-\infty}^{\infty} dk\;
{\theta(k)\over {\sqrt{2k}}}\;
[a(k,\tau)e^{-ikx}+{a^{\dag}}(k,\tau)e^{ikx}]\,\eqno(2.5)$$

\noindent where $a(k,\tau)$ and ${a^{\dag}}(k,\tau)$
satisfy
$[a(k,\tau),{a(k^\prime,\tau)}^{\dag}]=\delta(k-k^\prime)$ etc.
If we integrate  the momentum space expansion
of $\varphi'(x,\tau)$ over $x$
we  may show that $\,\varphi(\infty,\tau)-\varphi(-\infty,\tau)=0$.
Combining this with the condition obtained above we are led to
$\partial_{\tau}\varphi(\pm\infty,\tau)=0$ as another consistency condition.
This is similar to the condition $\partial_{t}\varphi(x^{1}=\pm\infty,t)=0$
which
in contrast is {\it added} to  the equal-time theory upon invoking
physical considerations.

The  following nonlocal
{\it constraint equation} in the {\it front form} theory then
follows when we
integrate the Lagrange eq. of motion (2.3) over the
(longitudinal) coordinate $x$ and use the constraint just obtained

$$\beta(\tau)\equiv \omega(\lambda\omega^2+{m_{0}}^2)+lim_{L\to\infty} {1\over
L}
 \int_{-L/2}^{L/2} dx \Bigl[ \,(3\lambda\omega^2+{m_{0}}^2)\varphi +
 \lambda (3\omega\varphi^2+\varphi^3 ) \,
\Bigr]=0.\,\eqno(2.6)
$$

\nl Eliminating  $\omega$ using (2.6)  would result in a  nonlocal
light-front
Hamiltonian [7] in the place of the local and polynomial type
assumed in the equal-time discussion.
In fact, {\it general arguments} which include the
{\it microcausality} postulate
may be given
on the light-front to show the possibility of the appearance of
nonlocality in the longitudinal coordinate,
while the theory remains local in the
transverse ones if present. We do find  in
ref. 1 some examples  where constraints on the
potential arise due to the necessity of incorporating special
relativity in the theory.
At the tree level,
since  $\varphi$ is
an ordinary bounded function, the second term in the expression of $\beta$
in (2.6) drops out and we obtain the conventional result $V'(\omega)=0$. In
the quantized theory (2.6) dictates the high order quantum
corrections which may lead to phase transition, say, from the symmetric to
an asymmetric phase. This problem has been
considered [18] recently in the context of the
discretized formulation (in finite volume)
where $\omega$  is an operator and its ordering
with nonzero modes, e.g. $\varphi$,
must  be taken care of when handling the constraint
equation and the renormalization of the theory. Here the commutator
of $\omega$ with $\varphi$  gives a  nonlocal and
non-linear expression [8,7]
in $\varphi$, which makes
the treatment difficult. On the other hand, in the continuum formulation
this commutator becomes vanishing and  $\omega$ is a c-number, which
makes the discussion of the renormalization below much simpler.
We remark in passing that the infinite volume limit must be
ultimately taken even if we work in the finite volume  to find genuine
physical effects.

\bigskip
\noindent {\bf 3- Renormalization. Phase transition  in two dimensions:}
\medskip

The theory based on (2.1) and (2.6) may be renormalized. We do not
solve (2.6) but instead obtain the
renormalized constraint equation.
We set   $M_{0}^{2}(\omega)= (m_{0}^{2}+3 \lambda \omega^2)$
and choose   ${\cal H}_0={M_0}^2\varphi^2/2$
so that  ${\cal H}_{int}= \lambda\omega\varphi^3+\lambda
\varphi^4/4$. In view of the  superrenormalizability of the two dimensional
theory we need to do only  the mass renormalization.
We assume that the bare mass is nonvanishing so that [19]
$\int dx \varphi(x,\tau)={\sqrt{2\pi}}\,\tilde\varphi(k=0,\tau)=0$, e.g.,
$k\equiv {k^{+}}>0$, and the corresponding term in (2.1) and (2.6)
drops out.
We could follow as is usually done [20] the old
fashioned perturbation theory, but the
Dyson-Wick expansion  based on the Wick theorem [17]

$$T[e^{i\int d^{2}x\, j(x)\varphi(x)}\,]=e^{-{1\over 2}\int\int d^{2}x
 d^{2}y \,j(x)G_{0}(x-y)j(y)}:\,[e^{i\int d^{2}x \,j(x)\varphi(x)}]:\,$$

\nl is convenient. Here $T$ indicates the ordering in $\tau$
and  $G_{0}$ is the free scalar field propagator.

The  self-energy correction  to the  {\it one loop order} is

$$\eqalignno{-i\Sigma(p)&=-i\Sigma_1-i\Sigma_2(p) \cr
&=(-i6\lambda){1\over 2} D_{1}({M_0}^2)+(-i6\lambda\omega)^2 {1\over 2}
(-i) D_2(p^2,{M_0}^2)\,,
&(3.1)\cr} $$

\noindent where the divergent contribution  $D_1$
refers to the one-loop tadpole while  $D_{2}$ to the
one-loop  finite contribution coming from the $\varphi^{3}$ vertex.
The latter carries the sign opposite to that of the first and it will be
argued below to be of the same order in $\lambda$
as the first one, because of the presence of $\omega$ in it. We have shown
explicitly the symmetry and other factors  in (3.1).
The one particle reducible graphs coming from the cubic vertex are
ignored and also $\langle  \varphi(x)\rangle=0$.
Following the straightforward procedure [21] and
using the dimensional regularization [22],
adopting the minimal subtraction  prescription, we obtain

$$\eqalignno{{D_1(M_0)&= {1\over {(2\pi)}^n}
\int {d^{n}k\over (k^2+{M_0}^2)} = \mu^{(n-2)}{1\over {4\pi}}
({{M_0}^2\over {4\pi\mu^2}})^{({n\over
2}-1)}\Gamma(1-{n\over2})}\cr &\to {\mu^{(n-2)}\over {4\pi}}\Bigl [{2\over
(2-n)}
-\gamma-ln({{M_0^2}\over {4\pi\mu^2}})\Bigr ],\,&(3.2)\cr}$$

\noindent where the limit   $n\to 2$ is to be taken at the end and we
suppress the terms which vanish in this limit.  Also

$$D_{2}(p^2,{M_0}^2)=\int {d^2k\over { (2\pi)^2}} { 1\over
{(k^2+M_0^2)[(p-k)^2+{M_0}^2]} },\qquad
D_2(p^2,M_{0}^2)\vert_{p^2=-M_{0}^2}=
{{\sqrt 3}\over {18M_{0}^2}}. \eqno(3.3) $$

The physical mass  $M(\omega)$ is defined [16,19] by

$${M_{0}}^2(\omega)+\Sigma(p)\vert_{p^2=-M^2(\omega)}
= M^2(\omega)\,\eqno(3.4)$$

\noindent where  $p^{\mu}$ is the Euclidean space 4-vector and
$M(\omega)$ determines the pole of the renormalized propagator. We
obtain from (3.1-4)

$$ {M_0}^{2}(\omega)=M^{2}(\omega)+{{3\lambda}\over
{4\pi}}\Bigl[\gamma+ln({M^2(\omega)\over {4\pi\mu^2}})
\Bigr]+18\lambda^2\omega^2 D_2(p,M^2)\vert_{p^2=-M^2}
+{3\lambda\over {2\pi}}{1\over {(n-2)}}.\,\eqno(3.5)$$

\noindent Here we have taken into account that in view of the tree level
result
$\,\omega(\lambda\omega^2+{m_0}^2)=\omega[{M_0}^2(\omega)-2\lambda\omega^2]
=0$ the correction term  $\lambda^2\omega^2 $ ( when $\omega \ne0$) is,
really  of the first order in
$\lambda$. We ignore terms of order $\lambda^{2}$ and higher
and  remind  that
$M_{0}$ depends on $\omega$ which in its turn is involved in the
constraint equation  (2.6).
To maintain consistency we  replace $M_0$ by $M$
in the terms that  are already  multiplied by  $\lambda$.


{}From  (3.5) we obtain the {\it mass renormalization condition }

$$M^2-m^2=3\lambda\omega^2+{{3\lambda}\over {4\pi}}
ln({m^2\over {M^2}})-\lambda^2
\omega^2 {\sqrt{3}\over {M^2}}\,\eqno(3.6)$$

\noindent where $M(\omega)\equiv M$ and
$M(\omega=0)\equiv m$ indicate the phsyical masses in the
asymmetric and symmetric phases respectively.
The eq. (3.6) expresses the invariance of the bare mass  and
for $\omega=0$ or $\lambda=0$ it implies $M^2=m^2$.

We next take the vacuum expectation value of the
constraint equation (2.6) in order to obtain another independent equation
which relates  $\omega$ with the other physical parameters of the theory.
To the lowest order [17] we find

$$\eqalignno{\quad{3\lambda\omega\langle\varphi(0)^2\rangle
&\simeq 3\lambda \omega.iG_{0}(x,x)}= 3\lambda\omega. D_1(M),\cr
{\lambda\langle\varphi(0)^3\rangle &\simeq \lambda
(-i{\lambda}\omega).6.\int dx
\langle T(\varphi(0)^{3}\varphi(x)^{3})\rangle_{c}^{0},}\cr
&= -6\lambda^2\omega D_3(M)=-6\lambda^2\omega {b\over{(4\pi)^2 M^2}}
,\,&(3.7)\cr}$$

\noindent where {\it c} indicates {\it connected} diagram and [17] $D_3$
is a finite integral like $D_{2}$ with three denominators and a numerical
computation gives $b\simeq 7/3$. From (2.6) on making use of (3.5-7) we
find that the divergent term cancels giving rise to the
{\it renormalized constraint equation}

$$\beta(\omega)\equiv\omega\Bigl[M^2-2\lambda\omega^2 +
\lambda^2\omega^2 {{\sqrt 3}\over M^2}
-{{6\lambda^2}\over (4\pi)^2}{b\over M^2}\Bigr]=0.\,\eqno(3.8)$$


We will verify below that $\beta$  coincides with
the total derivative with respect to $\omega$,
in the equal-time formulation, of the  (finite) difference $F(\omega)$ (see
Sec. 3 below)
of the  renormalized vacuum energy densities  in the {\it asymmetric}
($\omega\ne0$) and
{\it symmetric} ($\omega=0$) phases in the theory. The last term in $\beta$
corresponds to a correction $\simeq \lambda(\lambda\omega^2)$ in this energy
difference and thus may not be ignored just like in the case of the
self-energy discussed above.
In the equal-time case   (3.8) would be
required to be {\it added} to the theory
upon physical considerations. It will  ensure that
the sum of the tadpole diagrams, to the approximation concerned,
for the
transition $\varphi\to vacuum$ vanishes [22]. The physical outcome would then
be
the same in the two forms of treating the theory here  discussed.
The variational methods write only the first two ($\approx$ tree level)
terms in the  expression for $\beta$  and thus ignore
the terms coming from the finite corrections.
A similar remark can be made about the last term in (3.6).
Both of the eqs. (3.6) and (3.8) and
the difference of energy densities above are also found
to be independent of the arbitrary mass scale introduced in
the dimensional regularization and  contain only the finite
physical parameters of the theory.

Consider first the {\it symmetric phase} with $\omega\approx 0$,
which is allowed from  (3.8).
{}From (3.6) we  compute
$\partial M^2/\partial{\omega}=
2\lambda\omega(3-{\sqrt 3}\lambda/M^2)/[1+3\lambda/(4\pi M^2)-{\sqrt 3}
\lambda^2\omega^2/M^4]$ which is needed to find  $\beta'\equiv
d\beta/d\omega =d^2F/d\omega^{2}$, the second derivative of the above mentioned
energy
difference. Its  sign will determine the nature of the stability
of the vacuum.  We find
$\beta'(\omega=0)=M^2[1-0.0886(\lambda/M^2)^2]$, where by
the same arguments as made above in the case of $\beta$ we may not ignore
the $\lambda^{2}$ term. The $\beta'$
changes the sign from a positive value for vanishingly weak couplings to
a negative one  when the coupling increases.  In other words
the system starts out in a stable symmetric phase for very small coupling but
passes over into an unstable symmetric phase for values greater
than $g_{s}\equiv\lambda_{s}/(2\pi m^2)\simeq 0.5346$.


Consider next the case of {\it the spontaneously broken symmetry
phase} ($\omega\ne 0$). From (3.8) the values of $\omega$ are now given by

$$M^2-2\lambda\omega^2
+{{\sqrt 3}\lambda\over 2}=0,\,\qquad\qquad (\omega\ne0),\eqno(3.9)$$

\noindent where we have made use of the tree level approximation
$\, 2\lambda \omega^2\simeq M^2\,$ when $\omega\ne 0$.
The mass renormalization condition becomes

$$M^2-m^2=3\lambda\omega^2+{{3\lambda}\over {4\pi}}\,
ln({m^2\over {M^2}})-\lambda {\sqrt{3}\over2}.\eqno(3.10)$$

\noindent On eliminating $\omega$ from (3.9) and (3.10)
we obtain the {\it modified duality relation}

$${1\over {2}}M^2+m^2+{{3\lambda}\over {4\pi}}\,
ln({{m^2}\over {M^2}})+{{\sqrt 3}\over 4}\lambda=0.\,\eqno(3.11)$$

\noindent which can also be rewritten as
$[\lambda\omega^2+m^2+(3\lambda/({4\pi}))
ln(m^2/M^2)]=0$ and it shows that  the  real solutions exist only for
$M^2 > m^2$. The finite corrections found here are again
not considered in the references cited in Sec. 1, for
example, they  assume (or find)  the tree level expression
$\,M^{2}-2\lambda\omega^{2}=0$.
In terms of  the dimensionless coupling constants
$g=\lambda/(2\pi m^2)\ge 0$ and $G=\lambda/(2\pi M^2)\ge 0$
we  have  $G<g$.
The new self-duality eq. (3.11) differs from the old one [12,14] and
shifts  the critical coupling to a higher value.
We find that: {\it i)} for $g < g_c=6.1897$ ({\it critical coupling})
there is no real solution for $G$, {\it ii)} for a fixed $g>g_{c}$ we have two
solutions for $G$ one with the point lying on the upper branch
($G>1/3$) and the other with that on the lower branch ($
G<1/3$),  of the curve describing $G$ as a function of $g$ and
 which starts at the point $(g=g_{c}=6.1897,G=1/3)$,
{\it iii)} the lower branch with $G<1/3$,
approaches to a vanishing value for $G$ as $g\to\infty$, in contrast to the
upper one for which $1/3<G<g$ and  $G$ continues to increase.
{}From (3.10) and   $\beta=\omega[M^2-2\lambda\omega^2
+{{\sqrt 3}\lambda/2}]\,$ we  determine
$\beta'\approx (1+0.9405 G)$ which is always  positive and thus
indicates a minimum of the difference of the vacuum energy densities
for  the nonzero values of $\omega$.

The energetically favored broken symmetry phases
become available only after the
coupling grows to the critical coupling $g_c=6.18969$
and beyond this the asymmetric phases would be preferred against
the unstable symmetric phase in which the system finds itself when
$g>g_{s}\simeq 0.5346$. The phase transition is thus of the {\it second order}
confirming the conjecture of Simon-Griffiths. If we ignore the additional
finite renormalization corrections  we obtain
complete agreement with the earlier results, e.g.,
the symmetric phase always remains stable but for $g>1.4397$ the
energetically favored asymmetric phases also do appear, indicating a first
order transition.

\bigskip

\noindent {\bf Vacuum energy density:}
\medskip
The expression for the vacuum energy density in the equal-time
formulation is given by

$${\cal E(\omega)}=I_{1}(M_0)+{1\over 2}{m_0}^2\omega^{2}+\,{\lambda\over 4}
\omega^{4}+{{\lambda}\over 4}.3.{D_1(M_0)}^2+({-i6\lambda\omega})^{2}.{1\over
{2!}}.
 {1\over 6}.{D_3}(M_0).\eqno(3.12)$$

\nl Here the first term  is the vacuum energy density with respect to the free
propagator
with mass ${M_{0}}^{2}$ and is given by [12]

$$\eqalignno{{I_1(M_0)&={1\over (2\pi)^{(n-1)}}\int d^{(n-1)}k\,\,
{1\over 2}\,\,\sqrt{\vec k^{2}+{M_0}^2}
={{M_0}^{n}\over {(4\pi)^{n\over2}}}\,\,{1\over n}\,\,\Gamma(1-{n\over 2})}
\cr &\to {\mu^{(n-2)}}\,{{M_0^2}\over {4\pi}}\,{1\over 2}\,
\Bigl[\,{2\over {(2-n)}}+
1-\gamma-ln({{M_0}^2\over{4\pi\mu^2}})\,\Bigr]\,&(3.13)\cr}$$

\nl The ${D_{1}}^{2}$ term represents the two-loop correction of the order
$\lambda $  and so does  the last one in view of the
discussion above except for that it is finite  and
carries an opposite relative sign. We remark that the  last term
is non-vanishing even in the  light-front computation where we find
in the integrand  $\theta(k)\theta(k')\theta(k'')\delta(k+k'+k'')$
multiplied by another distribution. This product, however,
may not be considered vanishing. The last term of $\beta$ in (3.8) corresponds
to the
derivative with respect to $\omega$ of the last term in (3.12).

{}From (3.5) we have
$\,M_0^2\approx M^2 \Bigl[1+(3\lambda/{({2\pi}M^2)})
\{A+1/(n-2)\}\Bigr],\, $ where
$A=(4\pi)\Bigl[(1/{8\pi})ln(M^2/{\bar\mu}^2)+
3\lambda\omega^2 D_2\Bigr], \, {\bar\mu}^2=4\pi\mu^2exp(-\gamma),$ and
$\,\gamma\simeq 0.5772$. We rewrite (3.12) as

$$\Bigl[{M^n\over{(4\pi)^{n\over 2}}}\,{1\over
n}\Gamma(1-{n\over 2})\,-\,\mu^{(n-2)}{M^2\over {4\pi}}{1\over
(2-n)}\Bigr] $$

$$+{3\lambda\over{({4\pi})^2}}\Bigl[{M^{(n-2)}\over{(4\pi)^{{n\over 2}-1}}}\,
\Gamma(1-{n\over 2})\,-\mu^{(n-2)}{2\over
(2-n)}\Bigr]A $$

$$+{3\lambda\over{(4\pi)^2}}{\mu^{2(n-2)}}\Bigl[{{M^2}^{({n\over 2}-1)}
\over{(4\pi\mu^2)^{({n\over 2}-1)}}}\,
{1\over 2}\,\Gamma(1-{n\over 2})\,-{1\over
(2-n)} \Bigr]^2 $$

$$+{1\over 2}M^2\omega^2+{3\lambda\over{4\pi}}\omega^2\,A-{3\over
2}\lambda\omega^4+{\lambda\over 4}\omega^4$$

$$+(-i6\lambda\omega)^2.{1\over 6}.D_3(M)\,
+\mu^{(n-2)}{3\lambda\over{(4\pi)^2}}{1\over
(2-n)^2}+{m_0^2\over {4\pi}}{1\over
(2-n)}.\eqno(3.14)$$

\noindent Except for the last two terms containing the poles
the expression involves only the finite terms. Taking the limit   $n\to 2$
we obtain the following  finite expression
for the difference of the vacuum energy densities
in the broken  and the symmetric phases

$$\eqalignno{F(\omega)&={\cal E}(\omega)-{\cal E}(\omega=0)\cr
&={(M^2-m^2)\over {8\pi}}+{1\over {8\pi}}(m^2+3\lambda\omega^2)
\,ln({m^2\over M^2})+{3\lambda\over 4}\Bigl[{1\over {4\pi}}ln({m^2\over M^2})
\Bigr]^2\,\cr
&\qquad\quad+{1\over 2}m^2 \omega^2+{\lambda\over 4}\omega^4+{1\over {2!}}.
(-i6\lambda\omega)^2.{1\over 6}.D_3(M),\quad&(3.15)\cr} $$

\nl  which is also found to be
independent of the arbitrary mass $\mu $ on using  (3.6).

We verify  that $(dF/d\omega)=\beta$
and $d^2F/d\omega^{2}=\beta'$ and except for the last term
in (3.15) it coincides with the result in the earlier works.
{}From numerical computation we verify that at the minima
corresponding to the nonvanishing value of $\omega$
the value of F is negative
and that for a fixed $g$ it is more negative for the point on the
lower branch ($G<1/3$) than for that on the upper branch
($G>1/3$).  To illustrate  we find:
for $g=6.366$ and $ G=0.263$ we get $\vert\omega\vert
=0.736,\, F=-0.097\lambda $
while for the same $g$ but $G=0.431$ we find $\vert\omega\vert=0.617,
\, F=-0.082\lambda $. For $g=11.141$ and $ G=0.129$ we get $\vert
\omega\vert=1.050,\,
F=-0.174\lambda $
while for the same $g$ but $G=1.331$ we find $\vert\omega\vert=0.493,\,
F=-0.111\lambda $. The symbolic manipulation was found very handy in
treating the coupled eqs.  (3.6) and (3.8).
\bigskip
\nl {\bf 4- Conclusions:}
\medskip
The  present work and the earlier ones
on the mechanism of spontaneous continuous symmetry breaking  add
to the previous experience [2-6] that the {\it front form} dynamics is a useful
complementary method and needs to be studied systematically in the context of
QCD and other problems. The physical results following from
one or the other form of the theory
should come out to be the same though the mechanisms to arrive at them
may be  different. In the equal-time case
we are required to add external considerations in order
to constrain the theory (e.g., in the variational methods)
while  the analogous conditions in the light-front formulation
seem to be already contained in it through the  self-consistency equations.
When additional fields, e.g., fermionic ones are present also
the constraint equations in the theory quantized on the light-front
would relate the various types of vacuum condensates (vacuum expectation
values of composite scalar fields). The renormalized
QCD  on the light-front needs to be studied to see if it can
handle the difficulties of the conventional
formulation mentioned in the Introduction.
The phase transition of the second order  in the two dimensional $\phi^{4}$
theory follows if we include also the finite renormalization
corrections and without them our results agree with those obtained
in the variational methods. The conclusions here like in the  earlier
papers [12-18] are based on one loop order computation. However, we start
with the  lagrangian with the correct sign for the mass term and no
broken symmetry potential is invoked. At least
 close to the initial symmetric phase, corresponding to the
 small values of the coupling,
 the one loop order result should  be  reliable
approximation and it  is shown that the symmetric phase turns instable.
In this connection it would be interesting
to clarify the still undecided issue of  triviality in $3+1$
dimensions by
constructing  renormalized $\phi^{4}$ theory on the light-front.

\bigskip
\bigskip
\noindent{\bf Acknowledgements:}
\medskip

The author acknowledges
the hospitality offered to him by the {\sl Physics Department,
Ohio State University}, Columbus, Ohio,
and {\sl INFN and  Dipartimento di Fisica, Universit\`a di
Padova}, Padova. A research grant from INFN-Padova is gratefully
acknowledged.
Acknowledgements with thanks are due to Robert Perry and
Avaroth Harindranath
for constructive  clarifications and
discussions during the progress of the work,
to Ken Wilson for  asking useful questions and constructive suggestions,
to Stan Brodsky for encouragement,
to Mario Tonin, Stuart Raby, G. Costa,
A. Bassetto, and P. Marchetti  for constructive discussions.

\bigskip
\bigskip
\nl {\bf References:}
\medskip

\item{[1.]}P.A.M. Dirac, Rev. Mod. Phys. {\bf 21} (1949) 392.
\item{[2.]}S. Weinberg, Phys. Rev. {\bf 150} (1966) 1313.
\item{[3.]}J.B. Kogut and D.E. Soper, Phys. Rev. {\bf D 1} (1970) 2901.
\item{[4.]}S. Fubini, G. Furlan, and C. Rossetti, Nuovo Cimento {\bf 40}
(1965) 1171; H. Leutwyler, Springer Tracts Mod. Phys. {\bf 50} (1969) 29;
F. Rohrlich, Acta Phys. Austr. {\bf 32} (1970) 87.


\item{[5.]}K.G. Wilson, Nucl. Phys. B (proc. Suppl.) {\bf 17} (1990);
R.J. Perry, A. Harindranath, and K.G. Wilson,
Phys. Rev. Lett. {\bf 65} (1990)  2959.

\item{[6.]}S.J. Brodsky and H.C. Pauli, {\it Schladming Lectures}, {\sl SLAC}
 preprint {\sl SLAC}-PUB-5558/91.

\item{[7.]} P.P. Srivastava, {\it Higgs mechanism ({\sl tree level})
in light-front
quantized field theory},  Ohio State  preprint 92-0012,
 SLAC data base PPF-9202, December 91;
{\it Spontaneous symmetry breaking mechanism
in light-front quantized field theory}- {\sl Discretized formulation},
 Ohio State preprint  92-0173,  SLAC- PPF-9222, April 1992-
 {\sl The infinite volume or the continuum limit is also established}.
Papers contributed
to {\it XXVI Int. Conf. on High Energy Physics}, Dallas, Texas,
August 92, {\sl AIP Conference Proceedings 272, p. 2125, Ed. J.R. Sanford},
hep-th@xxx.lanl.gov/9412204 and /9412193;
{\it Constraints and Hamiltonian
in Light-front Quantized Theory},
Padova University preprint, DFPF/92/TH/58, December 92,
Nuovo Cimento {\bf A 107} (1993) 549.

\item{[8.]} The constraint was noted earlier in the two dimensional
discretized formulation of the scalar theory
:  Maskawa and K. Yamawaki, Prog. Theor. Phys. {\bf 56} (1976)
270; R.S. Wittman, in Nuclear and Particle Physics
on the Light-cone, eds.
M.B. Johnson and L.S. Kisslinger, World Scientific, Singapore, 1989.
In the context of the continuum formulation and in $3+1$ dimensions as well
it was discussed in [7].

\item{[9.]} P.A.M. Dirac, {\it Lectures
in Quantum Mechanics}, Benjamin, New York, 1964.

\item{[10.]} See for example,
D.J.F. Callaway, Phys. Rep. {\bf 167} (1988) 241.

\item{[11.]}B. Simon and R.B. Griffiths, Commun. Math. Phys. {\bf
33} (1973) 145;  B. Simon, {\it The ${P(\Phi)_{2}}$ Euclidean (Quantum)
Field Theory}, Princeton University Press, 1974.

\item{[12.]} See for example, M. Consoli and A. Ciancitto, Nucl. Phys.
{\bf B254} (1985) 653;  P.M. Stevenson, Phys. Rev. {\bf D 32} (1985) 1389;
S.J. Chang, Phys. Rev. {\bf D 13}
(1976) 2778.

\item{[13.]}G.V. Efimov, Int. Jl. Mod. Phys. {\bf A 4} (1989) 4977.
\item{[14.]}A. Harindranath and J.P. Vary, Phys. Rev. {\bf D 37}
(1985) 1389.

\item{[15.]}M. Funke, U. Kaulfuss, and H.Kummel,
Phys. Rev. {\bf D 35} (1987) 621.

\item{[16.]}L. Polley and U. Ritschel, Phys. Lett. {\bf B 221}(1989) 44.

\item{[17.]} C. Itzykson and J.B. Zuber, {\it Quantum
Field Theory}, McGraw-Hill, 1980;  G. Parisi, {\it Statistical Field Theory},
Addison-Wesley, 1988.

\item{18.]} C.M. Bender, S. Pinsky and B. van de Sande, Phys. Rev.
{\bf D48} (1993) 816 and OHSTPY-HEP-TH-93-014. See also
S. Huang and W. Lin, Ann. Phys. (NY) 226 (1993).

\item{[19.]} See also,
S. Schlieder and E. Seiler, Commun. Math. Phys. {\bf 25} (1972) 62.

\item{[20.]} See for example, A. Harindranath and R.J. Perry, Phys. Rev. {\bf D
43} (1991)
492.

\item{[21.]} The propagator for the free (e.g., in the interaction
representation)
massive scalar field is defined by
$iG_{0}(x;x')\equiv\langle 0\vert T\{\varphi(x,\tau)
\varphi(x',\tau')\}\vert 0\rangle$, where T indicates
the ordering in light-front time $\tau\equiv x^{+}$. From the light-front
momentum space expansion of the field $\varphi$ and the
commutation relations of the operators we arrive at
$iG_{0}(x;0)\equiv\langle 0\vert T(\varphi(x^{-},\tau)
\varphi(0,0))\vert 0\rangle
=$
\item{}$\int\,
{(dk^{+}/ {4\pi k^{+}})}\,\theta(k^{+})\,[\theta(\tau)\,
e^{-i(k^{+}x^{-}+\epsilon_k\tau)}+\theta(-\tau)\,e^{i(k^{+}x^{-}
+\epsilon_k\tau)}]$ where
$2k\,\epsilon_{k}=M_{0}^{2}$. Making use of the well known integral
representation of $\theta(\tau)$, simple manipulations, and the
identity $\,[\theta(k^{+})+\theta(-k^{+})]=1\,$,
true in the sense of distribution
theory, it may be rewritten as
$\int\int
(d{k^{+}}d{k^{-}}/{(2\pi)^2})
\,{[i (2k^{+}k^{-}-M_{0}^2+i\epsilon)^{-1}]}\,e^{-i(k^{+}x^{-}+k^{-}x^{+})}.$
Here  $ k^{\pm}$ are  now dummy
variables taking values from ${-\infty} $ to $\infty $
and the {\it integration over
$k^{-}$ is understood to be performed first}. This is  exactly
like what we find  in the case of equal-time quantization where
{\it integration  over the time component of the
momentum vector
is understood to be done first}, ( see also,
S.S. Schweber, {\it Relativistic Quantum Field Theory},
Row, Peterson and Co., NY, 1961, pg. 443).
In order to be able to make use of Euclidean space integrals in the context of
light-front quantization,
we make a change of variables
from $(k^{-},\,k^{+})$ to
$(k^{0}=(k^{-}+k^{+})/{\sqrt 2}, \;k^{+})$ so that the integration over
the {\it new} variable $k^{0}$, which also
varies from $-\infty$ to $\infty$,  is
now to be performed first. We may check that the locations of the
poles in the complex
$k^{0}$ plane corresponding to  $k^{+}>0$ and $k^{+}<0$ are such
that the (usual)
Wick rotation in the $k^{0}$ plane is permitted. We may also
make  another change of variables from
$(k^{0},\,k^{+})$ to $(k^{0},\;k^{1}=[{\sqrt2}k^{+}-k^{0}])$ which results in
$2 k^{+}k^{-}=({k^{0}}^{2}-{k^{1}}^{2}) $. Upon performing the Wick rotation
now in $k^{0}$ we obtain, for example,  the Euclidean space integral used
in (3.2).

\item{[22.]}J. Collins, {\it Renormalization}, Cambridge University Press,
1984.

\bye